\documentstyle[12pt,psfig]{article}
\let\section=\subsection     \let\subsection=\subsubsection                
\begin{document}
\hspace{9.5cm}MPG-VT-UR 99/97\\
\begin{center}
   {\large \bf MOTT MECHANISM AND ANOMALOUS CHARMONIUM SUPPRESSION}\\[5mm]
   K.~MARTINS and D.~BLASCHKE \\[5mm]
   {\small \it  Fachbereich Physik der Universit\"at Rostock \\
   Universit\"atsplatz 1, 18051 Rostock, Germany \\[8mm] }
\end{center}
\begin{abstract}\noindent
We investigate a scenario where conditions for {\sc Mott} dissociation of 
charmonium are fulfilled locally in ultrarelativistic heavy ion collisions 
at the short time scales before hadronization and study possible consequences 
to be drawn from observation of ``anomalous'' thresholds in the J/$\psi$ 
suppression pattern.
We suggest that for a compilation of experiments with different kinematics, 
variables presently in use to display the observed suppression
pattern should be rescaled by the $\gamma$ factor of the $c\bar c$ state
relative to the participant center of mass.
This systematics can be tested by inverse kinematics experiments as, 
e.g., Pb-S collisions and by sampling data sets in $p_T$ and $y$ of the 
dimuon pairs.
\end{abstract}

\section{Introduction}
Recently, an ``anomalous'' threshold behaviour in the centrality dependence of 
J/$\psi$ suppression has been observed by the NA50 experiment \cite{gonin} in 
Pb-Pb collisions at 158 A GeV/c. 
It has been speculated that this indication of a critical behaviour 
might signal quark-gluon plasma formation \cite{bo96,klns}
as suggested earlier by Matsui and Satz \cite{ms86} to be a consequence
of the {\sc Mott} mechanism for charmonium bound states in a plasma.

Properties of bound states are modified in dense matter.
Above critical densities $n_{\sc Mott}$ 
(temperatures T$_{\sc Mott}$) they undergo a {\sc Mott}
transition to unbound (but correlated) states in the continuum. 
For ground state hadrons as light quark bound states this transition
corresponds to the phase transition from hadronic to quark 
matter \cite{brrk}. 
For the deeply  bound J/$\psi$ at this transition there is still an
energetic threshold to be overcome either by kinetic hadron or parton impact
or by further compressing and heating the plasma \cite{kms,rbs,b91,mbq}. 
In this contribution, we want to investigate a scenario where conditions for 
{\sc Mott} dissociation of charmonium are 
fulfilled locally at the short time scales before hadronization sets in and 
study possible consequences for the NA50 experiment.

\renewcommand{\thefootnote}{\fnsymbol{footnote}}
\section[Mott dissociation of heavy quarkonia$^*$]
{Mott dissociation of heavy quarkonia
\footnote{The asterix indicates that the $Q\bar Q$ state is a color singlet but
not yet a state of the charmonium spectrum and will be denoted as 
{\it in statu nascendi}, see \protect\cite{bh}.}}
\renewcommand{\thefootnote}{\arabic{footnote}}
In order to determine the critical density 
$n_{\sc Mott}(Q\bar Q^*)$ at which the 
$Q\bar Q^*$ pair will not evolve into quarkonium but rather dissociate into 
a pair of heavy mesons, we employ here the heuristic criterion  
that at this density the nearest neighbor for a given heavy quark $Q$ is not 
a heavy antiquark $\bar Q$ but a light antiquark $\bar q$ of the medium 
\cite{rbs}. 
According to elaborated quantum mechanical approaches \cite{matsui,cugnon}, 
at short time scales $\tau < \tau_{\rm f}$  after the creation and color 
neutralisation, the heavy quark pair evolves like a diverging wave packet 
\begin{equation}
<r^2>_{Q\bar Q^*}(\tau) = <r^2>_{Q\bar Q} (\tau/\tau_{\rm f})^{\beta },
\end{equation}
until it reaches the asymptotic value $<r^2>_{Q\bar Q}$ of a quarkonium state 
in free space, we take $\beta = 2$ \cite{mbq,bo89}.
The mean squared distance of a heavy-light quark pair is
obtained from the nearest neighbor distribution function \cite{rbs} for a
plasma density which is proportional to the participant density 
$n_{\rm part}(t)$ measured in the c.m.s. proper time 
\begin{equation}
<r^2>_{Q\bar q^*}(\tau) = {\rm const}/n_{\rm part}(\tau \gamma)~,
\end{equation}    
where $\tau$ is the time in the rest system of the quarkonium$^*$ and 
the $\gamma$ factor 
\begin{equation}
\label{gamma}
\gamma=\sqrt{{\rm cosh}^2(\Delta y) + p_T^2/M^2_{\mu^+ \mu^-}}
\end{equation}
takes into account that the $Q \bar Q^*$ state is Lorentz boosted relative to 
the c.m.s. of the participants. 
The difference in the $\gamma$- factors of S-U and Pb-Pb collisions to be 
discussed in Sect. 3 is mainly due to different central 
rapidities of about $y_{c}=2.4$ and $y_{c}=3.0$, respectively, in the 
rapidity shift $\Delta y=y_{c}-y_{\mu^+\mu^-}$.

During the phase when 
$n_{\rm part}(t)$ rises to its maximum 
at $t_0 = \tau_0 \gamma \sim 1$ fm/c, 
it eventually exceeds the critical {\sc Mott} density  
$n_{\sc Mott}({Q\bar Q^*})$ for which 
$<r^2>_{Q\bar q^*}=<r^2>_{Q\bar Q^*}$ and instead of a quarkonium state 
a pair of heavy mesons will emerge from that region in the reaction plane for 
which 
\begin{equation}
\label{nmott}
n_{\rm part}({\bf b, s})>n_{\sc Mott}({Q\bar Q^*})
=\gamma^{\beta}n_{\sc Mott}({Q\bar Q})~,
\end{equation}
where 
$n_{\sc Mott}({Q\bar Q})={\rm const}~(\tau_{\rm f}/t_0)^\beta /<r^2>_{Q\bar Q}$
is the critical {\sc Mott} density for a ${Q\bar Q^*}$ pair created at rest in
the medium ($\gamma=1$). In Fig. 1, we demonstrate for $\beta=2$ that the 
occurence of ``anomalous'' behaviour ({\rm Mott} effect) in a given 
nucleus-nucleus collision depends on the kinematical conditions. 
The dependence on the scaled variable $n_{\rm part}/\gamma^{\beta}$,
however, is expected to uncover the threshold behaviour due to the 
{\rm Mott} dissociation. 
\begin{center}
\begin{minipage}{13cm}
\vspace{1cm}
\centerline{\psfig{figure=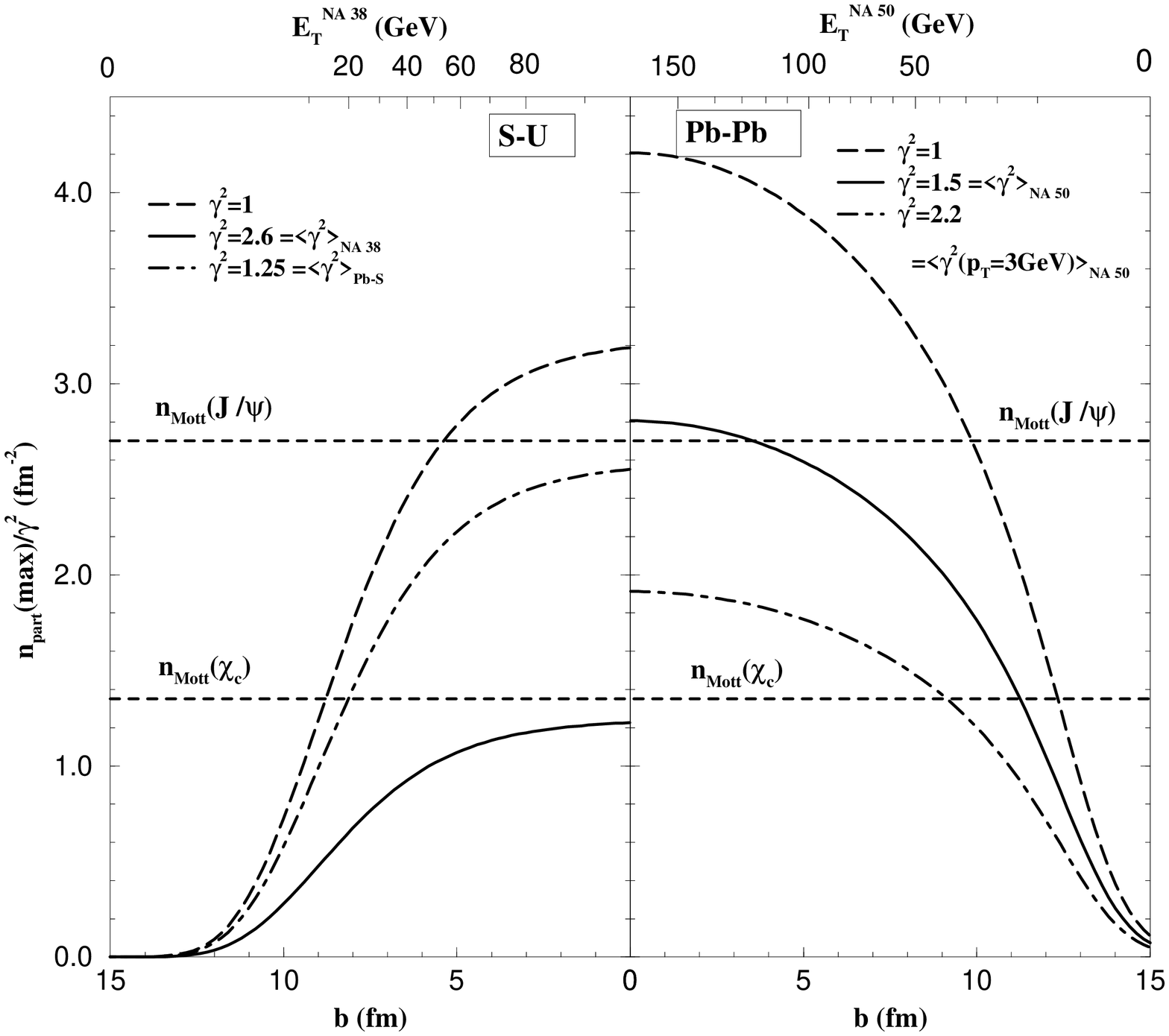,height=10cm,width=13cm,angle=-90}}
\baselineskip=12pt
\vspace{5mm}
{\begin{small}
Fig.~1. 
Criterion for Mott-dissociation of charmonium states.
Central participant densities (long dashed lines)
as a function of the impact parameter $b$ (bottom scale) or equivalently
the transverse energy $E_T$ (top scale) have to be
scaled by the $\gamma$- factor  
of the charmonium Lorentz-boost
relative to the c.m.s. of the collision, since the sizes  
of the charmonia {\it in statu nascendi} at $\tau_0$  depend on $\gamma$.
Solid lines correspond to the present data of the S-U (left panel) and the 
Pb-Pb (right panel) experiments, resp. 
The horizontal dashed lines show the critical {\sc Mott} densities which 
mark the onset of ``anomalous'' suppression as observed in the Pb-Pb 
experiment.
\end{small}}
\end{minipage}
\end{center}
\section{Glauber model for anomalous J/$\psi$ suppression}
Previous analyses of J/$\psi$ production in pA as well as AB collisions 
 within the Glauber model have shown \cite{klns,gh92} that nuclear absorption
 gives a satisfactory description
of J/$\psi$ suppression data except 
for the recent Pb-Pb data sample \cite{gonin} which therefore has been termed 
``anomalous''. The relevant quantity in this description is the survival 
probability 
$S_{\rm nucl}(b)=\int d^2 {\bf s}~S_{\rm nucl}({\bf b, s})$
for a J/$\psi$ formed in an AB collision at impact parameter $b=|{\bf b}|$ 
which is an integral over 
the survival probability density \cite{wong} 
\begin{equation}
\label{snucl}
S_{\rm nucl}({\bf b, s})=\frac{(1-(1-\sigma_{\rm abs}T_A({\bf s}))^A)(1-(1-\sigma_{\rm abs}T_B({\bf s-b}))^B)}
{\sigma_{\rm abs}^2~AB~ \int d^2 {\bf s}~T_A({\bf s})T_B({\bf s-b})}
\end{equation}
in the transverse plane.
$T_A$ ($T_B$) is the nuclear thickness function for a Woods-Saxon density 
distribution of the target (projectile) nucleus normalized to unity. 
A cross section $\sigma_{\rm abs}=7.3$ mb which is compatible with pA data 
gives an excellent description of J/$\psi$ production in AB collisions as 
performed by the NA38 experiment. 
For Pb-Pb collisions the nuclear absorption 
(\ref{snucl}) alone fails to describe the data, see Fig. 2.
Additional comover absorption \cite{gv96} does not give an overall satisfactory
description of the whole data sample even if a continuous increase of the 
absorption cross section with $n_{\rm part}$ is assumed, see also \cite{klns}. 

The rather pronounced threshold observed in the Pb-Pb 
data hints at a critical behaviour as, e.g., predicted by the {\sc Mott} 
dissociation effect considered in the previous Section.  
We want to implement this effect into the Glauber model approach by 
cutting those regions off the reaction plane 
where the {\sc Mott} condition (\ref{nmott}) is fulfilled. 
Considering a 30\% feeding of the observed J/$\psi$ from decays of 
$\chi_c$ and neglecting the small correction for $\psi'$ we obtain
the total J/$\psi$ survival probability
\begin{eqnarray}
\label{splasma}
{S(b)}&=& 0.7~ { S}_{\psi}(b)+0.3~{S}_{\chi}(b)
\nonumber\\
{S}_i(b)&=& \int d^2 {\bf s}~S_{\rm nucl}({\bf b, s})
~\Theta[n_{\rm part}({\bf b, s})-\gamma^\beta n^{\sc Mott}(i)]~.
\end{eqnarray}   
According to the Glauber model the density of participants in the 
transverse plane for a given impact parameter  is 
$n_{\rm part}({\bf b, s})= 
A T_A({\bf s})(1-(1-\sigma_{\rm inel} T_B({\bf s-b}))^B)
+B T_B({\bf s-b})(1-(1-\sigma_{\rm inel} T_A({\bf s}))^A)~$.
The total survival probability (\ref{splasma}) is a generalization
of equations employed in Refs. \cite{bo96,klns}. 
The main difference is that in Eq. (\ref{splasma}) we account for the fact 
that suppression of 
quarkonia {\it in statu nascendi} depends on the time scales involved. 
Consequently, the critical Mott-\-densities (\ref{nmott}) have to be rescaled 
with the $\gamma$-factor (\ref{gamma}) which implies a dependence on the 
rapidity shift $\Delta y$ and the transverse momentum $p_T$ at which the 
dimuon pair is detected.
In order to use different experimental situations for a more precise 
determination and identification of the thresholds, it is necessary to 
define a variable respective to which the threshold behaviour appears
universal. After inspection of Eq. (\ref{splasma}) we propose here the 
scaled variable 
$<n_{\rm part}(b)>/\gamma^\beta$~, where the mean number of participants
is defined \cite{klns} by
$<n_{\rm part}(b)>=
\int d^2 {\bf s}~ n_{\rm hard}({\bf b, s})~ n_{\rm part}({\bf b, s})/
\int d^2 {\bf s}~ n_{\rm hard}({\bf b, s})~$, 
and the density of hard collisions is 
$ n_{\rm hard}({\bf b, s})=A~B~T_A({\bf s})T_B({\bf s - b}) \sigma_\psi^{NN}$.
The numerical results are shown in Fig. 2.
\begin{center}
\begin{minipage}{13cm}
\centerline{\psfig{figure=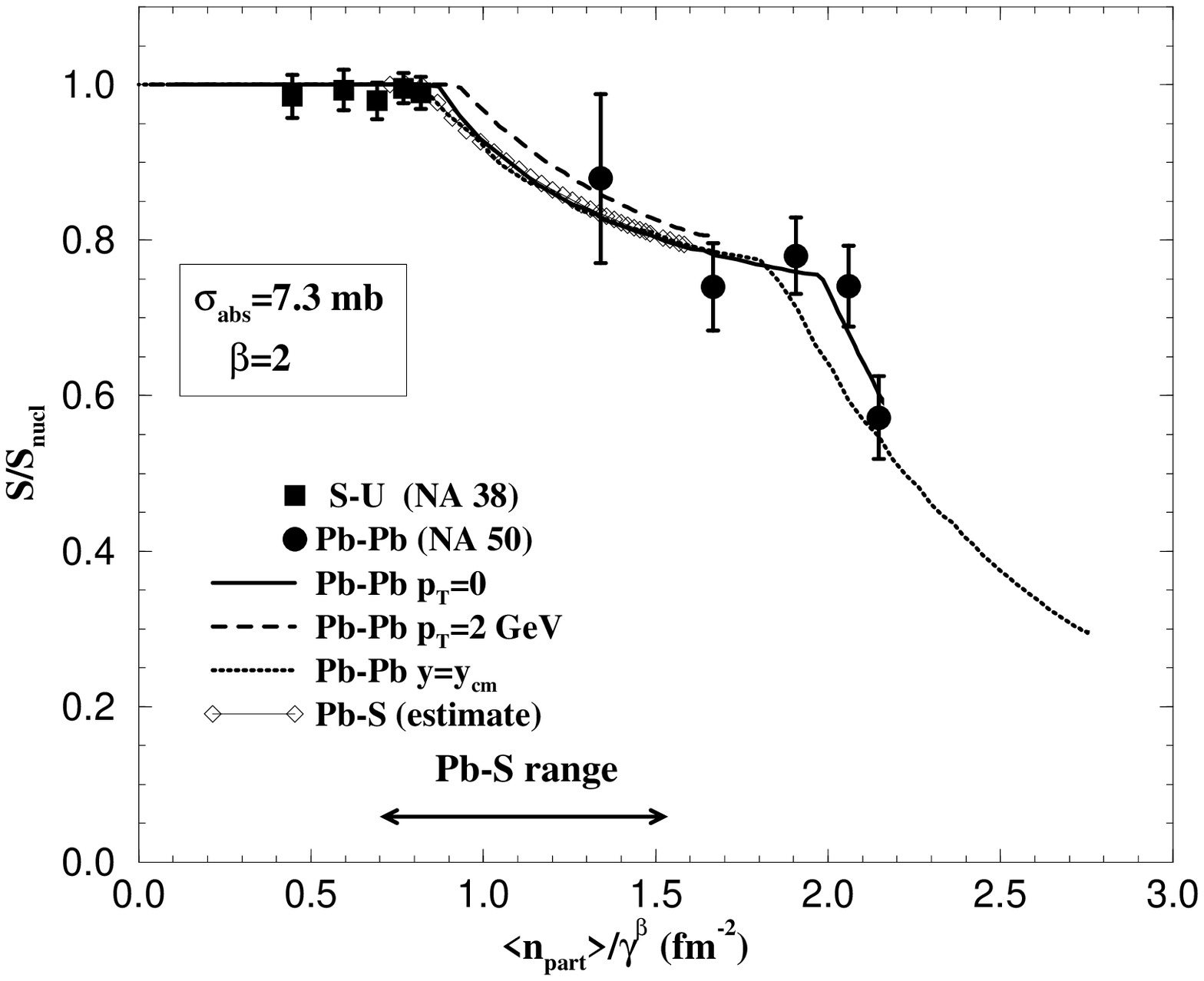,height=10cm,width=13cm,angle=0}}
\baselineskip=12pt
{\begin{small}
Fig.~2. J/$\psi$ suppression in nucleus-nucleus collisions
normalized to pure nucleonic absorption.
Whithin our approach, all data should approximately follow a universal curve 
as function of the averaged participant density divided by $\gamma^\beta$, 
where the two thresholds correspond to the onset of {\sc Mott} dissociation 
for the $\chi_c$ and the $\psi$ state, respectively.
The open diamonds show an estimate for the result of an inverse kinematics
experiment Pb-S which fills the gap between present S-U (filled squares) and 
Pb-Pb (filled circles) data sets.
\end{small}}
\end{minipage}
\end{center}
\section{Conclusions} 
We have reanalyzed recent data on J/$\psi$ production in S-U and Pb-Pb 
collisions at CERN-SpS and found that 
besides conventional sources an additional absorption mechanism with critical
behaviour is necessary. In the present work we have generalized the Glauber
model approach to J/$\psi$ suppression by implementing the critical {\sc Mott}
dissociation effect for charmonia {\it in statu nascendi}.

There are new aspects in this analysis: 
(1) there are two thresholds expected to occur in the ``anomalous''
J/$\psi$ suppression pattern due to the critical 
{\sc Mott} effect for the 1p ($\chi_c$) and the 1s (J/$\psi$) state;
(2) the corresponding threshold values for impact parameter or transverse 
energy depend on the kinematics of the $c \bar c$ pair relative to 
the hot and dense medium which it experiences {\it in statu nascendi};
(3) conclusions from J/$\psi$ suppression for the diagnostics of the state 
of matter (QGP formation) depend on the kinematics. 
As a consequence, QGP might be present in S-U collisions although J/$\psi$ 
production as measured by NA38 at $<\gamma^2>=2.6$ is not ``anomalous''.

In order to investigate the 
threshold behaviour of J/$\psi$ suppression
we have proposed the variable $<n_{\rm part}>/\gamma^\beta$,
where $\beta=2$ is supported by present data. 
We find it most useful to perform systematic experimental studies by
(1) using inverse kinematics (for lead beam on sulphur, see estimate in 
Fig. 2) and (2) sampling low vs. high $p_T$ and/or 
low vs. high rapidity bins.
These analyses can be performed using data as provided by the NA50 
experiment.

The theoretical approach to the kinetics of the 
hadronization process for a heavy quark pair in a dense plasma has been 
simplified here by the introduction of an
instantaneous breakup. A more detailed investigation is in 
progress \cite{diss}.

\end{document}